%% file: paper.tex
\documentclass{llncs}
\usepackage[utf8]{inputenc}
\usepackage{amsmath,amssymb,mathpazo}

\usepackage[colorlinks=true,allcolors=blue,breaklinks,draft=false]{hyperref} 
\usepackage{subcaption}
\usepackage[noend]{algpseudocode}
\usepackage{algorithm}
\usepackage[all]{xy}

\usepackage{verbatim}
\usepackage{booktabs}
\usepackage{enumitem}
\usepackage{etex}
\usepackage{tikz}

\usepackage{mathrsfs}
\usepackage{mathtools}
\usepackage{stmaryrd}

\include{macros}

\title{Faster linearizability checking via $P$-compositionality\thanks{This work is funded by a gift from Intel Corporation for research on Effective Validation of Firmware and the ERC project ERC 280053.}}
\author{Alex Horn \and Daniel Kroening}

\institute{University of Oxford}

\begin{document}

\maketitle


\begin{abstract}
Linearizability is a well-established consistency and correctness criterion for concurrent data types.
An important feature of linearizability is Herlihy and Wing's locality principle, which says that a concurrent system is linearizable if and only if all of its constituent parts (so-called objects) are linearizable.
This paper presents \emph{$P$-compositionality}, which generalizes the idea behind the locality principle to operations on the same concurrent data type.
We implement $P$-compositionality in a novel linearizability checker.
Our experiments with over nine implementations of concurrent sets, including Intel's TBB library, show that our linearizability checker is one order of magnitude faster and/or more space efficient than the state-of-the-art algorithm.
\end{abstract}

\section{Introduction}
\label{section:introduction}

\emph{Linearizability}~\cite{HW1990} is a well-established correctness criterion for concurrent data types and it corresponds to one of the three desirable properties of a distributed system, namely \emph{consistency}~\cite{GL2002}. The intuition behind linearizability is that every operation on a concurrent data type is guaranteed to take effect instantaneously at some point between its call and return.

The significance of linearizability for contemporary distributed key/value stores has been highlighted recently by the \emph{Jepsen} project, an extensive case study into the correctness of distributed systems.\footnote{\url{https://aphyr.com/posts/316-call-me-maybe-etcd-and-consul}} Interestingly, Jepsen found linearizability bugs in several distributed key/value stores despite the fact that they were designed based on formally verified distributed consensus protocols. This illustrates that there is often a gap between the design and the implementation of distributed systems. This gap motivates the study in this paper into runtime verification techniques (in the form of so-called \emph{linearizability checkers}) for finding linearizability bugs in a single run of a concurrent system.

The input to a linearizability checker consists of a sequential specification of a data type and a certain partially ordered set of operations, called a \emph{history}. A history represents a single terminating run of a concurrent system. We assume that the concurrent system is deadlock-free since there already exist good deadlock detection tools. Despite the restriction to single histories, the problem of checking linearizability is NP-complete~\cite{GK1997}. This high computational complexity means that writing an efficient linearizability checker is inherently difficult. The problem is to find ways of pruning a huge search space: in the worst case, its size is $O(N!)$ where $N$ is the length of the run of a concurrent system.

This paper presents a novel linearizability checker that efficiently prunes the search space by partitioning it into independent, faster to solve, subproblems. To achieve this, we propose \emph{$P$-compositionality} (Definition~\ref{def:compositionality}), a \emph{new partitioning scheme} of which Herlihy and Wing's locality principle~\cite{HW1990} is an instance. Recall that locality says that a concurrent system $Q$ is linearizable if and only if each concurrent object in $Q$ is linearizable. The crux of $P$-compositionality is that it generalizes the idea behind the locality principle to operations on the same concurrent object. For example, the operations on a concurrent unordered set and map are linearizable if and only if the \emph{restriction to each key} is linearizable. This is not a consequence of Herlihy and Wing's locality principle.

In this paper, we study the pragmatics of $P$-compositionality through its implementation in a novel linearizability checker and experimental evaluation. Our implementation is based on Wing~and~Gong's algorithm~(\emph{WG algorithm})~\cite{WG1993} and a recent extension by Lowe~\cite{L2015}. We call Lowe's extension of Wing and Gong's algorithm the \emph{WGL algorithm}. The idea behind the WGL algorithm is to prune states that are equivalent to an already seen state. Lowe's experiments show that the WGL algorithm can solve a significantly larger number of problem instances than the WG algorithm. We therefore use the more recent WGL algorithm as our starting point.

Our linearizability checker preserves three practical properties of the algorithms in the WG-family that we deem important. Firstly, our tool is precise, i.e., it reports no false alarms. This is particularly significant for evaluating large code bases, as effectively shown by the Jepsen project. Secondly, our tool takes as input an \emph{executable specification} of the data type to be checked. This significantly simplifies the task of expressing the expected behaviour of a data type because one merely writes code, i.e., no expertise in formal modeling is required. Finally, our tool can be easily integrated with a range of runtime monitors to generate a history from a run of a concurrent system. This is essential to make it a viable runtime verification technique.

We experimentally evaluate our linearizability checker using nine different implementations of concurrent sets, including Intel's TBB library, as exemplars of $P$-compositionality. Our experiments show that our linearizability checker is at least one order of magnitude faster and/or more space efficient than the WGL algorithm. Overall, the results of our work can therefore dramatically increase the number of runs that can be checked for linearizability bugs in a given time budget.

The rest of this paper is organized as follows. We first formalize the problem by recalling familiar concepts (\autoref{section:background}). We then present $P$-compositionality (\autoref{section:compositionality}) on which our decision procedure (\autoref{section:decision-procedure}) is based. We implement and experimentally evaluate our decision procedure (\autoref{section:experiments}). Finally, we discuss related work (\autoref{section:related-work}) and conclude the paper (\autoref{section:concl}).

\section{Background}
\label{section:background}

We recall familiar concepts that are fundamental to everything that follows.

\begin{definition}[History]
\label{def:history}
Let $E \deq \set{\mathsf{call}, \mathsf{ret}} \times \nats$. For all natural numbers $n$ in $\nats$, $\mathsf{call}_n \deq \pair{\mathsf{call}}{n}$ in $E$ is called a \defn{call} and $\mathsf{ret}_n \deq \pair{\mathsf{ret}}{n}$ in $E$ is called a \defn{return}. The invocation of a procedure with input and output arguments is called an \defn{operation}. An \defn{object} comprises a finite set of such operations. For all $e$ in $E$, $\mathit{obj}(e)$ and $\mathit{op}(e)$ denote the object and operation of $e$, respectively. A~\defn{history} is a tuple $\tuple{H, \mathit{obj}, \mathit{op}}$ where $H$ is a finite sequence of calls and returns, totally ordered by $\preceq_H$. When no ambiguity arises, we simply write $H$ for a history. We write $\abs{H}$ for the \defn{length} of $H$.
\end{definition}

Intuitively, a history $H$ records a particular run of a concurrent system. Using the implicitly associated functions $\mathit{obj}$ and $\mathit{op}$, a history $H$~gives relevant information on all operations performed at runtime, and the sequence of calls and returns in $H$ give the relative points in time at which an operation started and completed with respect to other operations. This can be visualized using the familiar history diagrams~\cite{HW1990}, as illustrated next.

\begin{figure}[t]
\begin{equation*}
\xymatrix@C=0.8em@R=0.5em{
  &   \mathsf{call}_1\ar@{|-|}[rrrr]^{\mathit{set}.\mathsf{insert(1) \colon \mathbf{true}}}    &&&&\mathsf{ret}_1& \\
  &&&&&&&&  \mathsf{call}_3\ar@{|-|}[rrrrrr]^{\mathsf{\mathit{set}.\mathsf{contains}(1) \colon \mathbf{true}}} &&&&&&\mathsf{ret}_3& \\
  && \mathsf{call}_2\ar@{|-|}[rrrrr]^{\mathit{set}.\mathsf{remove(1) \colon \mathbf{false}}}       &&&&&\mathsf{ret}_2&
}
\end{equation*}
\caption{A history diagram $H_1$ for the operations on a concurrent set}
\label{fig:history}
\end{figure}
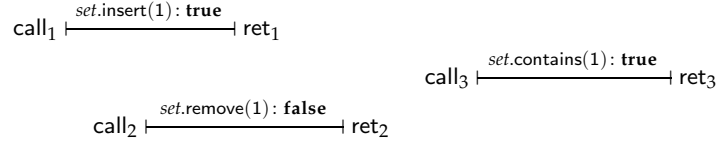

\begin{example}
Consider a concurrent set with the usual operations: `$\mathsf{insert}$' adds an element to a set, whereas `$\mathsf{remove}$' does the opposite, and `$\mathsf{contains}$' checks membership. The return value indicates the success of the operation. For example, `$\mathit{set}.\mathsf{remove}(1) \colon \mathbf{true}$' denotes the operation that successfully removed `$1$' from the object `$\mathit{set}$', whereas `$\mathit{set}.\mathsf{remove}(1) \colon \mathbf{false}$' denotes the operation that did not modify `$\mathit{set}$' because `$1$` is already not in the set. Then the history diagram in Fig.~\ref{fig:history} can be defined by $H_1 = \langle\mathsf{call}_1,\mathsf{call}_2,\mathsf{ret}_1,\mathsf{ret}_2,\mathsf{call}_3,\mathsf{ret}_3\rangle$ such that, for all $1 \leq i \leq 3$, $\mathit{obj}(\mathsf{call}_i) = \mathit{obj}(\mathsf{ret}_i) = \textrm`\mathit{set}\textrm'$, and the following holds:
\begin{itemize}
\item $\mathit{op}(\mathsf{call}_1) = \mathit{op}(\mathsf{ret}_1) = \textrm`\mathsf{insert(1) \colon \mathbf{true}}\textrm'$,
\item $\mathit{op}(\mathsf{call}_2) = \mathit{op}(\mathsf{ret}_2) = \textrm`\mathsf{remove(1) \colon \mathbf{false}}\textrm'$,
\item $\mathit{op}(\mathsf{call}_3) = \mathit{op}(\mathsf{ret}_3) = \textrm`\mathsf{contains(1) \colon \mathbf{true}}\textrm'$.
\end{itemize}
Note that $\abs{H_1} = 6$ and the total ordering $\preceq_{H_1}$ satisfies, among other constraints, $\mathsf{ret}_1 \preceq_{H_1} \mathsf{call}_3$ because $\mathsf{ret}_1$ precedes $\mathsf{call}_3$ in the sequence $H_1$.
\end{example}

Henceforth, we draw diagrams as in Fig.~\ref{fig:history}. Linearizability is ultimately defined in terms of sequential histories, in the following sense:

\begin{definition}[Complete and sequential history]
\label{def:sequential-history}
Let $e, e' \in E$ and $H$ be a history. If $e$ is a call and $e'$ is a return in $H$, both are \defn{matching} whenever $e \preceq_H e'$ and their objects and operations are equal, i.e.~$\mathit{obj}(e) = \mathit{obj}(e')$ and $\mathit{op}(e) = \mathit{op}(e')$. A history is called \defn{complete} if every call has a unique matching return. A complete history is called \defn{sequential} whenever it alternates between matching calls and returns (necessarily starting with a call).
\end{definition}

\begin{example}
\label{example:set-sequential-history}
The following history $H_2$ is sequential:
\begin{equation*}
\xymatrix@C=0.8em@R=1.2em{
& \ar@{|-|}[rrrrr]^{\mathsf{remove(1) \colon \mathbf{false}}} &&&&&  \ar@{|-|}[rrrrr]^{\mathsf{insert(1) \colon \mathbf{true}}} &&&&& \ar@{|-|}[rrrrr]^{\mathsf{contains(1) \colon \mathbf{true}}} &&&&&
}
\end{equation*}
And so is $H_3$ that we get when we swap the first two operations in $H_2$ (although the resulting sequence of operations is not what we would expect from a sequential set, as discussed next):
\begin{equation*}
\xymatrix@C=0.8em@R=1.2em{
& \ar@{|-|}[rrrrr]^{\mathsf{insert(1) \colon \mathbf{true}}} &&&&&  \ar@{|-|}[rrrrr]^{\mathsf{remove(1) \colon \mathbf{false}}} &&&&& \ar@{|-|}[rrrrr]^{\mathsf{contains(1) \colon \mathbf{true}}} &&&&&
}
\end{equation*}
\end{example}

$H_3$ in Example~\ref{example:set-sequential-history} illustrates that a history can be sequential even though it may not satisfy the expected sequential behaviour of the data type. This is addressed by the following definition:

\begin{definition}[Specification]
\label{def:specification}
A \defn{specification}, denoted by $\phi$ (possibly with a subscript), is a unary predicate on sequential histories.
\end{definition}

\begin{example}
\label{example:set-specification}
Define $\phi_\mathit{set}$ to be the specification of a sequential finite set. This means that, given a sequential history $S$ according to Definition~\ref{def:sequential-history}, the predicate $\phi_\mathit{set}(S)$ holds if and only if the input and output of  `insert', `remove' and `contains' in $S$ are consistent with the operations on a set. For example, $\phi_\mathit{set}(H_2) = \mathbf{true}$, whereas $\phi_\mathit{set}(H_3) = \mathbf{false}$ for the histories from Example~\ref{example:set-sequential-history}.
\end{example}

\begin{remark}
\label{remark:executable-specification}
In the upcoming decision procedure~(\autoref{section:decision-procedure}), every $\phi$ is an \emph{executable specification}. Informally, this is achieved by `replaying' all operations in a sequential history $S$ in the order in which they appear in $S$. If in any step the output deviates from the expected result, the executable specification returns false; otherwise, if it reaches the end of $S$, it returns true.
\end{remark}

The next definition will be key to answer which calls may be reordered in a history in order to satisfy a specification.

\begin{definition}[Happens-before]
\label{def:happens-before}
Given a history $H$, the \defn{happens-before} relation is defined to be a partial order $<_H$ over calls $e$ and $e'$ such that $e <_H e'$ whenever $e$'s matching return, denoted by $\mathsf{ret}(e)$, precedes $e'$ in $H$, i.e.~$\mathsf{ret}(e) \preceq_H e'$. We say that two calls $e$ and $e'$ \defn{happen concurrently} whenever $e \not<_H e'$ and $e' \not<_H e$.
\end{definition}

\begin{example}
\label{example:happens-before}
For the history $H_1$ in Fig.~\ref{fig:history}, we get:
\begin{itemize}
\item $\mathsf{call}_1 <_{H_1} \mathsf{call}_3$ and $\mathsf{call}_2 <_{H_1} \mathsf{call}_3$, i.e.~$\mathsf{call}_1$ and $\mathsf{call}_2$ happen-before $\mathsf{call}_3$;
\item $\mathsf{call}_1 \not<_{H_1} \mathsf{call}_2$ and $\mathsf{call}_2 \not<_{H_1} \mathsf{call}_1$, i.e.~$\mathsf{call}_1$ and $\mathsf{call}_2$ happen concurrently.
\end{itemize}
\end{example}

Note that a history $H$ is sequential if and only if $<_H$ is a total order. More generally, $<_H$ is an interval order~\cite{BEEH2105}: for every $x, y, u, v$ in $H$, if $x <_H y$ and $u <_H v$, then $x <_H v$ or $u <_H y$. Observe that a partial order $\pair{P}{\leq}$ is an interval order if and only if no restriction of $\pair{P}{\leq}$ is isomorphic to the following Hasse diagram~\cite{R1978}:
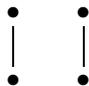
\begin{displaymath}
\xymatrix@C=1.5em@R=1.5em{
  \bullet           & \bullet            \\
  \bullet\ar@{-}[u] & \bullet\ar@{-}[u]
}
\end{displaymath}


Put differently, this paper is about a decision procedure (\autoref{section:decision-procedure}) that concerns a certain class of partial orders. The decision problem rests on the next definition:

\begin{definition}[Linearizability]
\label{def:linearizability}
Let $\phi$ be a specification. A \defn{$\phi$-sequential history} is a sequential history $H$ that satisfies $\phi(H)$. A history $H$ is \defn{linearizable with respect to $\phi$} if it can be extended to a complete history $H'$ (by appending zero or more returns) and there is a $\phi$-sequential history $S$ with the same $\mathit{obj}$ and $\mathit{op}$ functions as $H'$ such that
\begin{enumerate}[label=\textbf{L\arabic*}]
\item $H'$ and $S$ are equal when seen as two sets of calls and returns;\label{def:linearizability-L1}
\item $<_H\ \subseteq\ <_S$, i.e.~for all calls $e$, $e'$ in $H$, if $e$ happens-before $e'$, the same is true in $S$.\label{def:linearizability-L2}
\end{enumerate}
\end{definition}

Informally, extending $H$ to $H'$ means that all pending operations have completed. This paper therefore considers only complete histories. This is fully justified under our stated assumption (\autoref{section:introduction}) that the concurrent system is deadlock-free~\cite{L2015}. Condition~\ref{def:linearizability-L1} means that $H'$ and $S$ are identical if we disregard the order in which calls and returns occur in both sequences. Condition~\ref{def:linearizability-L2} says that the happens-before relation between calls in $H$ must be preserved in $S$.

\begin{example}
Recall Example~\ref{example:set-specification}. Then $H_1$ in Fig.~\ref{fig:history} is linearizable with respect to $\phi_\mathit{set}$ because $H_2$ is a witness for a $\phi_\mathit{set}$-sequential history that respects the happens-before relation $<_{H_1}$ detailed in Example~\ref{example:happens-before}. In particular, $\mathsf{call}_1 <_{H_1} \mathsf{call}_3$ and $\mathsf{call}_2 <_{H_1} \mathsf{call}_3$ cannot be reordered.
\end{example}

\section{$P$-compositionality}
\label{section:compositionality}

In this section, we introduce $P$-compositionality. We illustrate our new partitioning scheme in Examples~\ref{example:set-compositionality}--\ref{example:stack}.

\begin{definition}[$P$-compositionality]
\label{def:compositionality}
Let $P$ be a function that maps a history $H$ to a non-trivial partition of $H$, i.e. $P$ satisfies $P(H) \not= \set{H}$. A specification $\phi$ is called \defn{$P$-compositional} whenever any history $H$ is linearizable with respect to $\phi$ if and only if, for every history $H' \in P(H)$, $H'$ is linearizable with respect to $\phi$. When this equivalence holds we speak of \defn{$P$-compositionality}.
\end{definition}

In the following examples, we assume that the partitions are non-trivial. The first example illustrates that the locality principle~\cite{HW1990} is an instance of $P$-compositionality.

\begin{example}
Denote with $\mathit{Obj}$ the set of objects. Let $\phi$ be a specification for all objects in $\mathit{Obj}$. Let $P_\mathit{Obj}$ be the function that maps every history $H$ to the set of histories $\cH$ where each sub-history $H' \in \cH$ is the restriction of $H$ to an object in $\mathit{Obj}$. Then $P_\mathit{Obj}(H)$ is a partition of $H$. By the locality principle~\cite{HW1990}, a history $H$ is linearizable with respect to $\phi$ if and only if, for all $H_\mathit{obj} \in P_\mathit{Obj}(H)$, $H_\mathit{obj}$ is linearizable with respect to $\phi$. Therefore $\phi$ is a $P_\mathit{Obj}$-compositional specification.
\end{example}

The remaining examples show that $P$-compositionality strictly generalizes the locality principle because $P$-compositionality can partition a history even if the implementation details or constituent parts (i.e.~objects) of a concurrent system are unknown. For example, there are at least eight different implementations of concurrent sets (Table~\ref{table:mnemonics}), but we do not need to know the objects (e.g.~registers, buckets) of which such implementations consist in order to partition one of their histories. This is in contrast to the locality principle where such knowledge is required. Put differently, $P$-compositionality is all about the \emph{interface} of a concurrent data type, whereas the locality principle hinges on the \emph{implementation details} of such an interface.

\begin{example}
\label{example:set-compositionality}
Reconsider $\phi_\mathit{set}$, the specification of a set from Example~\ref{example:set-specification}, where all operations have the form $\mathsf{insert}(k)$, $\mathsf{remove}(k)$ and $\mathsf{contains}(k)$ for some $k$. Let $P_\mathit{set}$ be the function that partitions every history $H$ according to such $k$. Since the `insert', `remove' and `contains' operations on a single set object are linearizable if and only if the restriction to each $k$ is linearizable, $\phi_\mathit{set}$ is a $P_\mathit{set}$-compositional specification of a set.

Similarly, there exists a $P_\mathit{map}$-compositional specification for concurrent unordered maps where every history is partitioned by each key $k$.
\end{example}

\begin{example}
Consider a concurrent array. As their sequential counterparts, a concurrent array can be only read or written at a particular array index. Let $P_\mathit{array}$ be the function that partitions a history based on such array indexes. This gives a $P_\mathit{array}$-compositional specification of an array.
\end{example}

\begin{example}
\label{example:stack}
Consider a concurrent stack where each pop and push operation also returns the height of the stack before it is modified. Among other things, the return value can be used to determine whether the operation has succeeded. For example, if $\mathit{stack}.\mathsf{pop}$ returns zero, we know the pop operation was unsuccessful (and the popped element is undefined) because the stack was empty at the time the operation was called. We can use the returned height to partition a history such that a concurrent stack is linearizable if and only if each partition is linearizable. This way we get a $P_\mathit{stack}$-compositional specification of a stack.
\end{example}

Intuitively, the reason why the previous specifications are $P$-compositional is because all operations in one partition are, informally speaking, unaffected by all operations in every other partition. For example, the return value of $\mathit{set}.\mathsf{insert}(k)$ is unaffected by $\mathit{set}.\mathsf{insert}(k')$, $\mathit{set}.\mathsf{remove}(k')$ and $\mathit{set}.\mathsf{contains}(k')$ for $k \not= k'$. This clearly, however, has its limitations. For example, a `size' operation that returns the number of elements in a concurrent collection data type cannot be generally partitioned this way.

Note that all these examples have in common that their $P$-compositional specifications can be expressed as a conjunction of specifications that each partition a history. For example, $\phi_\mathit{set} = \bigwedge_{k \in K} \phi_{\mathit{set}(k)}$ where $\phi_{\mathit{set}(k)}$ for every $k$ is a sequential specification that only concerns operations on $k$, e.g.~$\mathit{set}.\mathsf{insert}(k)$.

Next, we show how to leverage the concept of $P$-compositionality to more efficiently find linearizability bugs.

\section{Decision procedure}
\label{section:decision-procedure}

In this section, we explain our linearizability checking algorithm that decides whether a history is linearizable with respect to some $P$-compositional specification (Definition~\ref{def:compositionality}). The novelty of our decision procedure is Algorithm~\ref{alg:create-sublogs} that leverages $P$-compositionality. In the next section (\autoref{section:experiments}), we experimentally evaluate the effectiveness of Algorithm~\ref{alg:create-sublogs}.

Since we base our work on the WGL algorithm (recall~\autoref{section:introduction}), we use the following data structures to represent the input to the decision procedure:
\begin{enumerate}
\item The specification (Definition~\ref{def:specification}) is modelled by a persistent data structure, e.g.~\cite{O1998}. Most standard data types in functional programming languages can be almost directly used this way. For instance, the specification of a set can be modelled through an immutable sequential set.
\item A history (Definition~\ref{def:history}), in turn, is represented by a doubly-linked list of so-called \defn{entries}. Consequently, each entry $e$ has a $e.\mathsf{next}$ and $e.\mathsf{prev}$ field that point to the next and previous entry, respectively. In addition, each entry $e$ has a $\mathsf{match}$ field, and we say that $e$ is a \defn{call entry} exactly if $e.\mathsf{match} \not= \mathbf{null}$; otherwise, $e$ is called a \defn{return entry}. Given a call entry $e$, $e.\mathsf{match}$ corresponds to the \defn{matching return entry} of $e$. This linked-list data structure therefore aligns directly with the usual definition of history (Definition~\ref{def:history}).
\end{enumerate}

The idea behind the WGL Algorithm~\ref{alg:linearizability-checker} is threefold: it keeps track of provisionally linearized call entries in a stack; it uses the stack to backtrack if necessary, and caches already seen configurations. We briefly explain each idea in turn. Denote the stack of call entries by $\mathsf{calls}$. Given a history $H$, the height of $\mathsf{calls}$ is at most half of $H$'s length, i.e.~$\abs{\mathsf{calls}} \leq 0.5 \times \abs{H} = N$. Note that there is no rounding involved because $\abs{H}$ is always even since every call entry has a matching return entry. The height of the stack grows only if a call entry can be linearized (line~\ref{line:is-linearizable}). When the stack grows or shrinks, the history is modified (lines~\ref{line:lift}~and~\ref{line:unlift}) by the $\Call{Lift}{}$ and $\Call{Unlift}{}$ procedures (Algorithm~\ref{alg:lift}). We remark that the workings of both procedures are illustrated by Example~\ref{example:lift}. If no further call entries can be linearized but the stack is nonempty, the algorithm backtracks and tries the next possible call entry (lines~\ref{line:backtrack-begin}--\ref{line:backtrack-end}). The backtracking points depend on the return value of $\mathit{apply}(\mathsf{entry}, \mathsf{s})$ and the cache. The former (line~\ref{line:apply}) models the specification $\phi$: by Remark~\ref{remark:executable-specification}, it determines whether $\mathsf{entry}$ can be applied to the current state $\mathsf{s}$ of a persistent data type. The latter (lines~\ref{line:cache-begin}--\ref{line:cache-end}) is an optimization due to Lowe~\cite{L2015} that prunes the search space by memoizing already seen configurations which are known to be non-linearizable. More accurately, each configuration is a pair that consists of a set of unique call entry identifiers and a state of the persistent data structure. The intuition behind pruning already seen configurations is that only one of two permutations of operations on a concurrent data type need to be considered if they lead to an identical state~\cite{L2015}. We remark that the total correctness of the WGL algorithm follows from Wing and Gong's total correctness argument~\cite{WG1993}.

\begin{algorithm}[t]
\caption{WGL linearizability checker~\cite{L2015}}
\begin{algorithmic}[1]
\Require $\mathsf{head\_entry}$ is such that $\mathsf{head\_entry}.\mathsf{next}$ points to the beginning of history $H$.
\Require $N = 0.5 \times \abs{H}$ is half of the total number of entries reachable from $\mathsf{head\_entry}$.
\Require $\mathsf{linearized}$ is a bitset (array of bits) such that $\mathsf{linearized}[k] = 0$ for all $0 \leq k < N$.
\Require For all entries $e$ in $H$, $0 \leq \mathit{entry\_id}(e) < N$.
\Require For all entries $e$ and $e'$ in $H$, if $\mathit{entry\_id}(e) = \mathit{entry\_id}(e')$, then $e = e'$.
\Require $\mathsf{cache}$ is an empty set and $\mathsf{calls}$ is an empty stack.
\While {$\mathsf{head\_entry}.\mathsf{next} \not= \mathbf{null}$}
\If {$\mathsf{entry}.\mathsf{match} \not= \mathbf{null}$} \Comment{Is call entry?}
  \State {$\pair{\mathsf{is\_linearizable}}{\mathsf{s}'} \gets \mathit{apply}(\mathsf{entry}, \mathsf{s})$} \Comment{Simulate entry's operation}\label{line:apply}
  \State {$\mathsf{cache'} \gets \mathsf{cache}$} \Comment{Copy set}\label{line:cache-begin}
  \If {$\mathsf{is\_linearizable}$}\label{line:is-linearizable}
    \State {$\mathsf{linearized'} \gets \mathsf{linearized}$} \Comment{Copy bitset}
    \State {$\mathsf{linearized'[\mathit{entry\_id}(\mathsf{entry})]} \gets 1$} \Comment{Insert $\mathit{entry\_id}(\mathsf{entry})$ into bitset}
    \State {$\mathsf{cache} \gets \mathsf{cache} \cup \set{\pair{\mathsf{linearized'}}{\mathsf{s}'}}$} \Comment{Update configuration cache}\label{line:cache-configuration}
  \EndIf\label{line:cache-end}
  \If {$\mathsf{cache'} \not= \mathsf{cache}$}
    \State {$\mathsf{calls} \gets \mathit{push}(\mathsf{calls}, \pair{\mathsf{entry}}{\mathsf{s}})$} \Comment{Provisionally linearize call entry and state}
    \State {$\mathsf{s} \gets \mathsf{s}'$} \Comment{Update state of persistent data type}
    \State {$\mathsf{linearized[\mathit{entry\_id}(\mathsf{entry})]} \gets 1$} \Comment{Keep track of linearized entries}
    \State {$\Call{Lift}{\mathsf{entry}$}}\label{line:lift} \Comment{Provisionally remove the entry from the history}
    \State {$\mathsf{entry} \gets \mathsf{head\_entry}.\mathsf{next}$} \Comment{Continue search in shortened history}
  \Else \Comment{Cannot linearize call entry}
    \State {$\mathsf{entry} \gets \mathsf{entry}.\mathsf{next}$} \Comment{Continue search in unmodified history}
  \EndIf
\Else \Comment{Handle ``return entry''}
  \If {$\mathit{is\_empty}(\mathsf{calls})$}\label{line:backtrack-begin}
    \State {\Return $\mathbf{false}$} \Comment{Cannot linearize entries in history}
  \EndIf
  \State {$\pair{\mathsf{entry}}{\mathsf{s}} \gets \mathit{top}(\mathsf{calls})$} \Comment{Revert to earlier state}
  \State {$\mathsf{linearized[\mathit{entry\_id}(\mathsf{entry})]} \gets 0$}
  \State {$\mathsf{calls} \gets \mathit{pop}(\mathsf{calls})$}
  \State {$\Call{Unlift}{\mathsf{entry}}$}\label{line:unlift} \Comment{Undo provisional linearization}
  \State {$\mathsf{entry} \gets \mathsf{entry}.\mathsf{next}$}\label{line:backtrack-end}
\EndIf
\EndWhile
\State {\Return $\mathbf{true}$}
\end{algorithmic}
\label{alg:linearizability-checker}
\end{algorithm}

\begin{figure}[b]
\centering
\subcaptionbox{\label{subfig:log}}{
\xymatrix@C=0.5em@R=1em{
  \mathsf{call}_1\ar@{|-|}[rrrr]^{\mathit{set}.\mathsf{insert(0) \colon \mathbf{true}}}      &&&& \mathsf{ret}_1   &&& \\
  & \mathsf{call}_2\ar@{|-|}[rrrrr]^{\mathit{set}.\mathsf{contains(0) \colon \mathbf{true}}} &&&&& \mathsf{ret}_2     \\
  && \mathsf{call}_3\ar@{|-|}[rrrrr]^{\mathit{set}.\mathsf{remove(1) \colon \mathbf{false}}} &&&&& \mathsf{ret}_3
}
}
\qquad
\subcaptionbox{\label{subfig:log-after-lift}}{
\xymatrix@C=0.4em@R=1em{
  \mathsf{call}_1\ar@{|-|}[rrrrr]^{\mathit{set}.\mathsf{insert(0) \colon \mathbf{true}}}      &&&&& \mathsf{ret}_1 \\
  && \mathsf{call}_3\ar@{|-|}[rrrrrr]^{\mathit{set}.\mathsf{remove(1) \colon \mathbf{false}}} &&&&&& \mathsf{ret}_3
}
}
\caption{
After calling $\textsc{Lift}(\mathsf{call}_2)$ in history~\eqref{subfig:log}, we get the history in~\eqref{subfig:log-after-lift}. $\textsc{Unlift}(\mathsf{call}_2)$ reverts this change in constant-time.
}
\label{fig:log}
\end{figure}
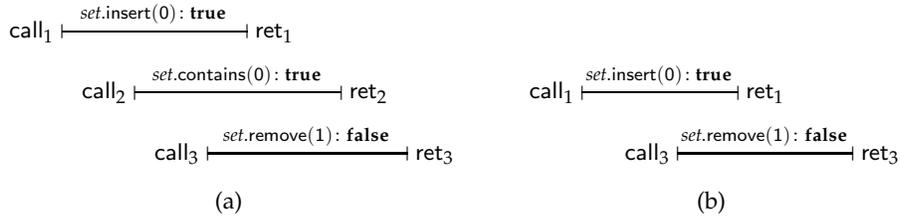

\begin{example}
\label{example:lift}
We illustrate the handling of entries in the history data structure. For this, consider the two histories in Fig.~\ref{fig:log}. In Fig.~\ref{subfig:log}, the entries satisfy the following: $\mathsf{call}_2.\mathsf{prev} = \mathsf{call}_1$, $\mathsf{call}_2.\mathsf{next} = \mathsf{call}_3$ and $\mathsf{call}_2.\mathsf{match} = \mathsf{ret}_2$ etc. Then $\textsc{Lift}(\mathsf{call}_2)$ (Algorithm~\ref{alg:lift}) produces the history shown in Fig.~\ref{subfig:log-after-lift}. Note that both $\mathsf{call}_2$ and $\mathsf{ret}_2$ are still valid entry pointers whose fields remain unchanged. This explains how $\textsc{Unlift}(\mathsf{call}_2)$ reverts the change in constant-time.
\end{example}

Algorithm~\ref{alg:create-sublogs} gives our partitioning scheme. This is an iterative algorithm that, given an entry in a history $H$ and positive integer $n$, partitions $H$ starting from that entry into at most $n$ separate sub-histories. The partitioning is controlled by the function $\mathit{partition} \colon E \to \nats$ from the set of call and return entries to the natural numbers.

\begin{example}
Consider the history in Fig.~\ref{subfig:log-after-lift}. For all entries $e$ in this history, let $\mathit{partition}(e) = k$ where $k$ is the integer argument of the operation. For example, $\mathit{partition}(\mathsf{call}_3) = \mathit{partition}(\mathsf{ret}_3) = 1$ because $\mathit{op}(\mathsf{call}_3) = \mathit{op}(\mathsf{ret}_3) = \textrm`\mathsf{remove(1) \colon \mathbf{false}}\textrm'$. Then the function $\Call{Partition}{\mathsf{call}_1}$ returns two disjoint sub-histories for the operations on `$0$' and `$1$', respectively:
\begin{equation*}
\xymatrix@C=0.8em@R=1em{
  \mathsf{call}_1\ar@{|-|}[rrrr]^{\mathit{set}.\mathsf{insert(0) \colon \mathbf{true}}}      &&&& \mathsf{ret}_1 \\
  & \mathsf{call}_2\ar@{|-|}[rrrrr]^{\mathit{set}.\mathsf{contains(0) \colon \mathbf{true}}} &&&&& \mathsf{ret}_2     \\
}
\qquad
\text{and}
\qquad
\xymatrix@C=0.8em@R=1em{
  \mathsf{call}_3\ar@{|-|}[rrrrrr]^{\mathit{set}.\mathsf{remove(1) \colon \mathbf{false}}} &&&&&& \mathsf{ret}_3.
}
\end{equation*}
\end{example}

Given a nonempty set of disjoint sub-histories returned by the $\Call{Partition}{}$ function (Algorithm~\ref{alg:create-sublogs}), we invoke Algorithm~\ref{alg:linearizability-checker} on each sub-history. It is not too difficult to implement sub-histories such that there is no sharing between them, and Algorithm~\ref{alg:linearizability-checker} could be therefore run in parallel for each sub-history. Nevertheless, this addresses a challenging problem that was identified independently by Lowe~\cite{L2015} and Kingsbury~\cite{K2014}.

\begin{theorem}
Let $\phi$ be a $P$-compositional specification and $H$ be a history. Denote with $\mathsf{head\_entry}$ the entry that represents the beginning of $H$. Associate with each disjoint history $H_k$ in partition $P(H)$ a unique number $0 \leq k < |P(H)| = n$. If, for all $H_k \in P(H)$ and $e \in H_k$, $\mathit{partition}(e) = k$, then $H$ is linearizable with respect to $\phi$ if and only if Algorithm~\ref{alg:linearizability-checker} returns true for every history in $\Call{Partition}{\mathsf{head\_entry}, n}$.
\end{theorem}

\begin{figure*}[t]
  \centering
\begin{minipage}{0.47\textwidth}
\vspace{0pt}
\begin{algorithm}[H]
\caption{History modifications}
\begin{algorithmic}[1]
\Procedure{Lift}{$\mathsf{entry}$}
  \State {$\mathsf{entry}.\mathsf{prev}.\mathsf{next} \gets \mathsf{entry}.\mathsf{next}$}
  \State {$\mathsf{entry}.\mathsf{next}.\mathsf{prev} \gets \mathsf{entry}.\mathsf{prev}$}
  \State {$\mathsf{match} \gets \mathsf{entry}.\mathsf{match}$}
  \State {$\mathsf{match}.\mathsf{prev}.\mathsf{next} \gets \mathsf{match}.\mathsf{next}$}
  \If {$\mathsf{match}.\mathsf{next} \not= \mathbf{null}$}
    \State {$\mathsf{match}.\mathsf{next}.\mathsf{prev} \gets \mathsf{match}.\mathsf{prev}$}
  \EndIf
\EndProcedure
\State
\Procedure{Unlift}{$\mathsf{entry}$}
  \State {$\mathsf{match} \gets \mathsf{entry}.\mathsf{match}$}
  \State {$\mathsf{match}.\mathsf{prev}.\mathsf{next} \gets \mathsf{match}$}
  \If {$\mathsf{match}.\mathsf{next} \not= \mathbf{null}$}
    \State {$\mathsf{match}.\mathsf{next}.\mathsf{prev} \gets \mathsf{match}$}
  \EndIf
  \State {$\mathsf{entry}.\mathsf{prev}.\mathsf{next} \gets \mathsf{entry}$}
  \State {$\mathsf{entry}.\mathsf{next}.\mathsf{prev} \gets \mathsf{entry}$}
\vspace{0.133em}
\EndProcedure
\end{algorithmic}
\label{alg:lift}
\end{algorithm}
\end{minipage}\hfill
\begin{minipage}{0.47\textwidth}
\vspace{0pt}
\begin{algorithm}[H]
\caption{History partitioner}
\begin{algorithmic}[1]
\Require {$n$ is a positive integer}
\Require {$\mathsf{entries}$ is an array of size $n$}
\Function{Partition}{$\mathsf{entry}$, $n$}
  \For {$0 \leq i < n$}
    \State {$\mathsf{entries}[i] \gets \mathbf{null}$}
  \EndFor
  \While {$\mathsf{entry} \not= \mathbf{null}$}
    \State {$i \gets \mathit{partition}(\mathsf{entry})\ \mathbf{mod}\ n$}
    \If {$\mathsf{entries}[i] \not= \mathbf{null}$}
      \State {$\mathsf{entries}[i].\mathsf{next} \gets \mathsf{entry}$}
    \EndIf
    \State {$\mathsf{next\_entry} \gets \mathsf{entry}.\mathsf{next}$}
    \State {$\mathsf{entry}.\mathsf{prev} \gets \mathsf{entries}[i]$}
    \State {$\mathsf{entry}.\mathsf{next} \gets \mathbf{null}$}
    \State {$\mathsf{entries}[i] \gets \mathsf{entry}$}
    \State {$\mathsf{entry} \gets \mathsf{next\_entry}$}
  \EndWhile
  \State \Return {$\mathsf{entries}$}
\EndFunction
\end{algorithmic}
\label{alg:create-sublogs}
\end{algorithm}
\end{minipage}
\vspace{-0.1em}
\end{figure*}

We next experimentally quantify the benefits of the previous theorem.

\section{Implementation and experiments}
\label{section:experiments}

In this section, we discuss and experimentally evaluate our implementation of the decision procedure (\autoref{section:experiments}). As an exemplar of $P$-compositionality, our experiments use Intel's TBB library and Lowe's implementations of concurrent sets. 

\subsection{Implementation}
\label{subsection:implementation}

The implementation details of an NP-complete decision procedure matter, especially for our experimental evaluation of $P$-compositionality. We particularly consider hashing and cache eviction options because these were not studied in previous implementations of the WG-based algorithms~\cite{WG1993,L2015}.

For experimental robustness, we implemented our linearizability checker in C++11~\cite{CPP11} because this language has built-in concurrency support while allowing us to rule out interference from managed runtime environments (e.g. JVM) due to garbage collection etc. The choice of language, though, meant that we had to implement persistent data structures from scratch. In doing so, we focused on optimizing equality checks for our specific purposes. This way, we managed to avoid a known performance bottleneck in Lowe's implementation of the WGL algorithm~\cite{L2015} where the cost of equality checks had to be compensated with an additional union-find data structure. Another optimization in our implementation is a constant-time (instead of linear-time) hash function for bitsets where we exploit the fact that the bitwise XOR operator over fixed-size bit vectors forms an abelian group. This optimization turns out to be important when histories are longer than 8\,K, cf.~\cite{L2015}. To see this, consider the computational steps for retrieving a configuration from the cache and updating it (line~\ref{line:cache-configuration} in Algorithm~\ref{alg:linearizability-checker}). For example, a history of length $2^{16}$ means that each bitset in a configuration is at least 3\,KiB, and so a constant-time hash function can make a measurable difference when the cache is frequently accessed. In fact, it is not uncommon for the cache to contain more than 27\,K of such configurations. For this reason, we also implemented a \emph{least recently used} (LRU) cache eviction feature that can optionally be enabled at compile-time. The effects of the LRU cache will be evaluated shortly.

Overall, our implementation and experimental setup is around 4\,K lines of code, including several dozen unit tests. All the code and benchmarks are publicly available in our source code repository.\footnote{\url{https://github.com/ahorn/linearizability-checker}}

\subsection{TBB and concurrent set experiments}
\label{subsection:set-experiments}

For the experimental evaluation of our partitioning scheme, we collected over $700$ histories from nine different implementations of concurrent sets by Lowe~\cite{L2015} and the concurrent unordered set implementation in Intel's TBB library.\footnote{\url{https://www.threadingbuildingblocks.org/}} We performed all experiments on a 64-bit machine running GNU/Linux 3.17 with $12$ Intel Xeon 2.4\,GHz cores and 94\,GB of main memory.

Each history is generated by running 4 concurrent threads that pseudo randomly invoke operations on a single shared concurrent set. The argument of each operation is a pseudo random uniformly distributed integer between $0$ (inclusive) and $24$ (exclusive). Each thread invokes 70\,K such operations. Note that this is significantly more than in previous experiments where each process is limited to $2^{13} \approx \text{8\,K}$ operations~\cite{L2015}. In total, since every call generates a pair of entries, every history $H$ in our benchmarks has length $\abs{H} = 4 \times 2 \times \text{70\,K} = \text{560\,K}$. We discuss the experimental results using Intel's TBB library and Lowe's concurrent set implementations in turn.

\begin{table}[t]
\centering
\begin{tabular}{@{}l||r|r|r||r|r|r||r|r|r@{}}
\toprule
&
\multicolumn{3}{|c||}{\textbf{WGL}}
&
\multicolumn{3}{|c||}{\textbf{WGL+LRU}}
&
\multicolumn{3}{|c}{\textbf{WGL+P}}   \\
Benchmark                                            & Time      & \multicolumn{1}{|c|}{Memory}
                                                                              & Timeout      & Time     & Memory            & Timeout  & Time          & Memory            & Timeout      \\ \midrule
TBB                                                  & 101\,s    & 9792\,MiB  & 0\%          & 11\,s    & \textbf{670\,MiB} & 0\%      & \textbf{6\,s} & 672\,MiB          & 0\%          \\ \midrule
CRLSL                                                & 20\,s     & 15738\,MiB & 0\%          & 25\,s    & 678\,MiB          & 0\%      & \textbf{6\,s} & \textbf{400\,MiB} & 0\%          \\ \midrule
CRLFSL                                               & 14\,s     & 15029\,MiB & 0\%          & 18\,s    & 678\,MiB          & 0\%      & \textbf{5\,s} & \textbf{401\,MiB} & 0\%          \\ \midrule
FGL                                                  & 16\,s     & 14297\,MiB & 0\%          & 81\,s    & 678\,MiB          & 0\%      & \textbf{5\,s} & \textbf{401\,MiB} & 0\%          \\ \midrule
LLL                                                  & 23\,s     & 16494\,MiB & 0\%          & 94\,s    & 678\,MiB          & 0\%      & \textbf{6\,s} & \textbf{401\,MiB} & 0\%          \\ \midrule
LSL                                                  & 20\,s     & 15736\,MiB & \textbf{0\%} & 25\,s    & 678\,MiB          & 14\%     & \textbf{6\,s} & \textbf{401\,MiB} & \textbf{0\%} \\ \midrule
LFLL                                                 & 11\,s     & 11847\,MiB & 0\%          & 15\,s    & 678\,MiB          & 0\%      & \textbf{5\,s} & \textbf{402\,MiB} & 0\%          \\ \midrule
LFSL                                                 & 14\,s     & 14712\,MiB & 0\%          & 18\,s    & 678\,MiB          & 0\%      & \textbf{5\,s} & \textbf{401\,MiB} & 0\%          \\ \midrule
LFSLF0                                               & 14\,s     & 13125\,MiB & 0\%          & 18\,s    & 678\,MiB          & 0\%      & \textbf{5\,s} & \textbf{402\,MiB} & 0\%          \\ \midrule
LFSLF1                                               & $<\,$1\,s & 404\,MiB   & 0\%          & $<\,$1\,s& 407\,MiB          & 0\%      & $<\,$1\,s     & 402\,MiB          & 0\%          \\ \midrule
OPTIMIST                                             & 16\,s     & 13818\,MiB & \textbf{0\%} & 54\,s    & 678\,MiB          & 9\%      & \textbf{5\,s} & \textbf{401\,MiB} & \textbf{0\%} \\
\bottomrule
\end{tabular}\vspace*{0.1cm}
\caption{Experimental results for three variants of the same linearizability checker. The results for the baseline are reported in the WGL column. The rows correspond to benchmarks drawn from Intel's TBB library and Lowe's implementations of concurrent sets (see Table~\ref{table:mnemonics} for mnemonics).}
\label{table:experiments}
\vspace{-1.5em}
\end{table}

The experimental results are given in Table~\ref{table:experiments}. Each of the three main columns corresponds to one variant of the same linearizability checker: `WGL' is the baseline, `WGL+LRU' is the WGL algorithm with LRU cache eviction enabled (\autoref{subsection:implementation}), and `WGL+P' is the WGL algorithm combined with our partitioning algorithm (Algorithm~\ref{alg:create-sublogs} in \autoref{section:decision-procedure}). We tried to use the WG algorithm~\cite{WG1993} without the extension by Lowe~\cite{L2015} but WG times out on the majority of benchmarks. We therefore do not report the results on the WG algorithm and focus on WGL, WGL+LRU and WGL+P. The meaning of the sub-columns is as follows. The `Time' and `Memory' columns give the average of the elapsed time and virtual memory usage, respectively. These averages exclude runs that we had to terminate after $1$ hour. The percentage of such terminated runs is given in the `Timeout' column.  In each row, all variants are compared with respect to the same benchmark data. We therefore do not report confidence intervals.

The TBB benchmark corresponds to the first row in Table~\ref{table:experiments} and consists of a total of $100$ histories. Table~\ref{table:experiments} clearly shows that the WGL+P algorithm is at least one order of magnitude faster compared to the baseline. We also see that enabling the LRU cache eviction decreases the memory footprint by at least one order of magnitude, approximately 10\,GiB versus 700\,MiB. In fact, the runtime performance of WGL+LRU is almost one order of magnitude faster than the baseline. The WGL+P algorithm is at least as fast and almost as space efficient as WGL+LRU. In the experiments with Lowe's implementations of concurrent sets (see next paragraph), we further investigate the effect of the LRU cache eviction feature and how it compares to the partitioning scheme.

We give Lowe's implementations of concurrent sets mnemonics (Table~\ref{table:mnemonics}) that identify the remaining ten benchmarks in Table~\ref{table:experiments}. Each of these ten benchmarks comprises between 50 and 100 histories with an average of $70$ histories per benchmark. To avoid bias, we collected these using Lowe's tool. The significance of the experimental results in Table~\ref{table:experiments} is twofold. Firstly, they show that on average, WGL+P is three times faster than WGL, and WGL+P consumes one order of magnitude less space than WGL. Secondly, and more crucially, however, these experiments reveal that WGL+LRU is not as efficient as WGL+P, in neither time nor space. For example, for WGL+LRU the average elapsed time of the FGL and LLL benchmark is 81\,s and 94\,s, respectively, with an average memory usage of 678\,MiB in both cases. By contrast, WGL+P achieves an average runtime of less than 7\,s (and so WGL+P is one order of magnitude faster than WGL+LRU) and consumes even less memory on average (401\,MiB) than WGL+LRU. The higher average runtime of WGL+LRU in the FGL benchmark is due to a single check that took several orders of magnitude longer (3068\,s) than the remaining checks (20\,s on average when the 3068\,s outlier is excluded). In the LLL benchmark there are two such outliers (2201\,s and 675\,s, whereas the other checks average 27\,s). The observed difference between WGL+LRU and WGL+P is even more pronounced in both the LSL and OPTIMIST benchmarks where the LRU cache eviction causes $14\%$ and $9\%$ of runs to timeout, whereas the WGL+P algorithm always runs to completion in less than a few seconds.

This experimentally confirms that the WGL+P is one order of magnitude faster as well as more space efficient than the baseline and WGL+P consumes even less space than our WGL+LRU implementation.

\begin{table}[t]
\centering
\begin{tabular}{l|l||l|l}
\toprule
Benchmark name                                                     & Mnemonic & Benchmark name                                                    & Mnemonic \\ \midrule
\scriptsize{collision resistance lazy skip list}                   & CRLSL    & \scriptsize{lock-free linked-list}                                & LFLL     \\
\scriptsize{collision resistance lock-free skip list}              & CRLFSL   & \scriptsize{lock-free skip list}                                  & LFSL     \\
 \scriptsize{fine-grained lock}                                    & FGL      & \scriptsize{lock-free skip list faulty (bad hash)}                & LFSLF0   \\
 \scriptsize{lazy linked-list}                                     & LLL      & \scriptsize{lock-free skip list faulty (good hash)}               & LFSLF1   \\
 \scriptsize{lazy skip list}                                       & LSL      & \scriptsize{optimistic lock}                                      & OPTIMIST \\
\bottomrule
\end{tabular}\vspace*{0.1cm}
\caption{Mnemonics for Lowe's implementation of concurrent sets~\cite{L2015}}
\label{table:mnemonics}
\vspace{-2.7em}
\end{table}

\vspace{-0.3em}
\section{Related work}
\vspace{-0.3em}
\label{section:related-work}

Linearizability is related to the concept of atomicity, including weaker forms such as $k$-atomicity~\cite{AAB2005}. An important difference is that atomicity is typically not defined in terms of a sequential specification, e.g.~\cite{WS2005}. The theoretical limitations of automatically verifying linearizability are well understood. Of course, the problem is generally undecidable~\cite{BEEH2013}. In fact, even checking finite-state implementation against atomic specifications, provided the number of program threads is bounded, is \textsc{EXPSPACE}~\cite{AMP2000}. And the best known lower bound for this problem is \textsc{PSPACE}-hardness. This explains the restrictions in this paper and its focus on runtime verification instead.

The literature on machine-assisted techniques for checking linearizability can be broadly divided into simulation-based methods (e.g.~\cite{CDG2005,DSW2011}), model checking (e.g.~\cite{VYY2009,BDT2010,CRZCA2010,LCLSZD2013}), static analysis (e.g.~\cite{ARRSY2007,BLMRS2008,V2009,V2010}) and fully automatic testing (e.g.~\cite{WG1993,ALSTW2010,SBASVY2011,FLR2011,PG2012,PG2013,GHL2013,L2015}). The simulation-based methods have been used by experts to mechanically verify simple fine-grained and lock-free implementations. Model checking requires less expertise but is typically limited to very small programs and a small number of threads due to the state explosion problem. By contrast, static analysis tools aim to prove correctness with respect to an unbounded number of threads. In general, these techniques are necessarily incomplete and require the user to supply linearization points and/or invariants. Vafeiadis~\cite{V2010} proposes a more automatic form of static analysis that works well on simpler concurrent data types such as stacks but reportedly not so well on data types that have more complicated invariants, including the CAS-based and lazy concurrent sets extensively studied in our experiments.

Our work is most closely related to linearizability testing techniques that are precise, fully automatic and necessarily incomplete, e.g.~\cite{WG1993,ALSTW2010,SBASVY2011,FLR2011,PG2012,PG2013,GHL2013,L2015}. We focus our discussion on tools that do not require the notion of commit points, cf.~\cite{ETQ2005}. The work in~\cite{ALSTW2010,GHL2013} checks $k$-atomicity with a polynomial-time algorithm assuming that each write to a register assigns a distinct value. By contrast, we solve a more general NP-complete problem of which $k$-atomicity is an instance. The tool in~\cite{SBASVY2011} analyzes code that uses concurrent collection data types such as maps. To make the analysis scale, the authors assume that the collection data types are linearizable, whereas our tool could be used to check such an assumption. A different tool~\cite{FLR2011} requires programmers to annotate concurrent implementations with so-called state summary functions that act as a form of specification. Our approach is more modular because it strictly separates the concurrent implementation from its specification. By contrast,~\cite{PG2012} works without the programmer having to provide a sequential specification. As a result, however, the tool can only find linearizability violations when an exception is thrown or a deadlock occurs. Subsequent work~\cite{PG2013} circumvents this, in the context of object-oriented programs, by considering the special case of a superclass serving as an executable, possibly non-deterministic, specification for all its subclasses. The fact that the superclass can be non-deterministic may explain why even checks of two threads can take a significant amount of time (e.g. 108\,min) despite the fact that each concurrent test considers only two possible linearizations~\cite{PG2013}. By contrast, the WGL algorithm~\cite{WG1993,L2015}, on which our decision procedure is based (\autoref{section:decision-procedure}), is significantly faster but limited to deterministic specifications. Crucially, our experiments (\autoref{section:experiments}) with $P$-compositional specifications show a significant improvement over the WGL algorithm.

\section{Concluding remarks}
\vspace{-0.3em}
\label{section:concl}

We have presented a precise, fully automatic runtime verification technique for finding linearizability bugs in implementations of concurrent data types that are expected to satisfy a $P$-compositional specification. Our experiments show that our partitioning scheme improves the WGL algorithm~\cite{WG1993,L2015} by one order of magnitude, in both time and space. An additional strength of our technique is that it is applicable to any linearizability checker. For this, however, our work assumes that the specification is $P$-compositional. This is generally not always the case and it would be therefore interesting to further generalize $P$-compositionality, perhaps with a less modular partitioning scheme that can make more assumptions about the underlying decision procedure.

\vspace{-0.3em}
\paragraph{Acknowledgements.} We would like to thank Gavin Lowe, Kyle Kingsbury and Alexey Gotsman for invaluable discussions.

\bibliographystyle{splncs}
\vspace{-0.8em}
\bibliography{paper}

\end{document}

%% file: macros.tex
\newcommand{\pair}[2]{\mbox{$\langle #1,\; #2 \rangle$}}
\newcommand{\tuple}[1]{\mbox{$\langle #1 \rangle$}}

\newcommand{\set}[1]{\mbox{$\{ #1 \}$}}

\newcommand{\deq}{\triangleq}

\newcommand{\nats}{\mathbb{N}}

\newcommand{\cH}{\mathcal{H}}

\newcommand{\defn}[1]{\textbf{#1}}

\DeclarePairedDelimiter\abs{\lvert}{\rvert}

%% file: paper.bbl
\begin{thebibliography}{10}

\bibitem{HW1990}
Herlihy, M.P., Wing, J.M.:
\newblock Linearizability: A correctness condition for concurrent objects.
\newblock ACM Trans. Program. Lang. Syst. \textbf{12}(3) (July 1990)  463--492

\bibitem{GL2002}
Gilbert, S., Lynch, N.:
\newblock Brewer's conjecture and the feasibility of consistent, available,
  partition-tolerant web services.
\newblock SIGACT News \textbf{33}(2) (June 2002)  51--59

\bibitem{GK1997}
Gibbons, P.B., Korach, E.:
\newblock Testing shared memories.
\newblock SIAM J. Comput. \textbf{26}(4) (August 1997)  1208--1244

\bibitem{WG1993}
Wing, J.M., Gong, C.:
\newblock Testing and verifying concurrent objects.
\newblock J. Parallel Distrib. Comput. \textbf{17}(1-2) (January 1993)
  164--182

\bibitem{L2015}
Lowe, G.:
\newblock Testing for linearizability.
\newblock In: PODC'15. (2015) Under submission.
  \url{http://www.cs.ox.ac.uk/people/gavin.lowe/LinearizabiltyTesting/}.

\bibitem{BEEH2105}
Bouajjani, A., Emmi, M., Enea, C., Hamza, J.:
\newblock Tractable refinement checking for concurrent objects.
\newblock In: POPL'15, ACM (2015)  651--662

\bibitem{R1978}
Rabinovitch, I.:
\newblock The dimension of semiorders.
\newblock Journal of Combinatorial Theory, Series A \textbf{25}(1) (1978)  50
  -- 61

\bibitem{O1998}
Okasaki, C.:
\newblock Purely Functional Data Structures.
\newblock Cambridge University Press (1998)

\bibitem{K2014}
Kingsbury, K.:
\newblock Computational techniques in {Knossos}.
\newblock \url{https://aphyr.com/posts/314-computational-techniques-in-knossos}
  (May 2014)

\bibitem{CPP11}
{ISO}:
\newblock International Standard ISO/IEC 14882:2011(E) Programming Language
  C++.
\newblock International Organization for Standardization (2011)

\bibitem{AAB2005}
Aiyer, A., Alvisi, L., Bazzi, R.A.:
\newblock On the availability of non-strict quorum systems.
\newblock In: Proceedings of the 19th International Conference on Distributed
  Computing. DISC'05, Springer (2005)  48--62

\bibitem{WS2005}
Wang, L., Stoller, S.D.:
\newblock Static analysis of atomicity for programs with non-blocking
  synchronization.
\newblock In: PPoPP '05, ACM (2005)  61--71

\bibitem{BEEH2013}
Bouajjani, A., Emmi, M., Enea, C., Hamza, J.:
\newblock Verifying concurrent programs against sequential specifications.
\newblock In: ESOP'13, Springer (2013)  290--309

\bibitem{AMP2000}
Alur, R., McMillan, K., Peled, D.:
\newblock Model-checking of correctness conditions for concurrent objects.
\newblock Inf. Comput. \textbf{160}(1-2) (July 2000)  167--188

\bibitem{CDG2005}
Colvin, R., Doherty, S., Groves, L.:
\newblock Verifying concurrent data structures by simulation.
\newblock Electron. Notes Theor. Comput. Sci. \textbf{137}(2) (2005)  93--110

\bibitem{DSW2011}
Derrick, J., Schellhorn, G., Wehrheim, H.:
\newblock Mechanically verified proof obligations for linearizability.
\newblock ACM Trans. Program. Lang. Syst. \textbf{33}(1) (January 2011)
  4:1--4:43

\bibitem{VYY2009}
Vechev, M., Yahav, E., Yorsh, G.:
\newblock Experience with model checking linearizability.
\newblock In: SPIN'09, Springer (2009)  261--278

\bibitem{BDT2010}
Burckhardt, S., Dern, C., Musuvathi, M., Tan, R.:
\newblock Line-up: A complete and automatic linearizability checker.
\newblock SIGPLAN Not. \textbf{45}(6) (June 2010)  330--340

\bibitem{CRZCA2010}
\v{C}ern\'{y}, P., Radhakrishna, A., Zufferey, D., Chaudhuri, S., Alur, R.:
\newblock Model checking of linearizability of concurrent list implementations.
\newblock In: CAV'10. (2010)  465--479

\bibitem{LCLSZD2013}
Liu, Y., Chen, W., Liu, Y.A., Sun, J., Zhang, S.J., Dong, J.S.:
\newblock Verifying linearizability via optimized refinement checking.
\newblock IEEE Trans. Softw. Eng. \textbf{39}(7) (2013)  1018--1039

\bibitem{ARRSY2007}
Amit, D., Rinetzky, N., Reps, T., Sagiv, M., Yahav, E.:
\newblock Comparison under abstraction for verifying linearizability.
\newblock In: CAV'07, Springer (2007)  477--490

\bibitem{BLMRS2008}
Berdine, J., Lev-Ami, T., Manevich, R., Ramalingam, G., Sagiv, M.:
\newblock Thread quantification for concurrent shape analysis.
\newblock In: CAV '08, Springer (2008)  399--413

\bibitem{V2009}
Vafeiadis, V.:
\newblock Shape-value abstraction for verifying linearizability.
\newblock In: VMCAI '09, Springer (2009)  335--348

\bibitem{V2010}
Vafeiadis, V.:
\newblock Automatically proving linearizability.
\newblock In: CAV'10, Springer (2010)  450--464

\bibitem{ALSTW2010}
Anderson, E., Li, X., Shah, M.A., Tucek, J., Wylie, J.J.:
\newblock What consistency does your key-value store actually provide?
\newblock HotDep'10, USENIX Association (2010)  1--16

\bibitem{SBASVY2011}
Shacham, O., Bronson, N., Aiken, A., Sagiv, M., Vechev, M., Yahav, E.:
\newblock Testing atomicity of composed concurrent operations.
\newblock SIGPLAN Not. \textbf{46}(10) (October 2011)  51--64

\bibitem{FLR2011}
Fonseca, P., Li, C., Rodrigues, R.:
\newblock Finding complex concurrency bugs in large multi-threaded
  applications.
\newblock In: EuroSys '11, ACM (2011)  215--228

\bibitem{PG2012}
Pradel, M., Gross, T.R.:
\newblock Fully automatic and precise detection of thread safety violations.
\newblock SIGPLAN Not. \textbf{47}(6) (June 2012)  521--530

\bibitem{PG2013}
Pradel, M., Gross, T.R.:
\newblock Automatic testing of sequential and concurrent substitutability.
\newblock In: ICSE'13, IEEE Press (2013)  282--291

\bibitem{GHL2013}
Golab, W., Hurwitz, J., Li, X.S.:
\newblock On the k-atomicity-verification problem.
\newblock ICDCS '13, IEEE Computer Society (2013)  591--600

\bibitem{ETQ2005}
Elmas, T., Tasiran, S., Qadeer, S.:
\newblock {VYRD}: Verifying concurrent programs by runtime refinement-violation
  detection.
\newblock SIGPLAN Not. \textbf{40}(6) (June 2005)  27--37

\end{thebibliography}
